\documentclass[preprint,showpacs,prb]{revtex4-1}%
\usepackage{amsfonts}
\usepackage{amsmath}
\usepackage{amssymb}
\usepackage{graphicx}
\usepackage{txfonts}%
\providecommand{\U}[1]{\protect\rule{.1in}{.1in}}

\begin{document}
\title{Optically controlled phase gate for two spin qubits in coupled quantum dots}
\author{Li-Bo Chen$^{1,2}$}
\author{L. J. Sham$^{2}$}
\email{lsham@ucsd.edu}
\author{Edo Waks$^{3}$}
\affiliation{$^{1}$Department of Physics, Ocean University of China, Qingdao 266100,
People's Republic of China}
\affiliation{$^{2}$Center for Advanced Nanoscience, Department of Physics, University
of California--San Diego, La Jolla, California 92093-0319}
\affiliation{$^{3}$Joint Quantum Institute, University of Maryland and National Institute
of Standards and Technology, College Park, Maryland 20742, USA }
\date{\today}

\begin{abstract}
We present a feasible scheme for performing an optically controlled phase gate
between two conduction electron spin qubits in adjacent self assembled quantum
dots. Interaction between the dots is mediated by the tunneling of the valence
hole state which is activated only by applying a laser pulse of the right
polarization and frequency. Combining the hole tunneling with the Pauli
blocking effect, we obtain conditional dynamics for the two quantum dots,
which is the essence of our gating operations. Our results are of explicit
relevance to the recent generation of vertically stacked self-assembled InAs
quantum dots, and show that by a design which avoids unintended dynamics the
gate could be implemented in theory in the 10 ps range and with a fidelity
over 90\%. Our proposal therefore offers an accessible path to the
demonstration of ultrafast quantum logic in quantum dots.

\end{abstract}

\maketitle

\section{Introduction}

Self-assembled semiconductor quantum dots (SAQDs) possess many properties
similar to real atoms, while simultaneously providing highly tunable
properties for controlling and manipulating individual electron
spins.\cite{Petroff,Bayer} Such SAQDs are promising candidates as qubits for
quantum computation, owing in part to the long coherence
time,\cite{Birkedal,Borri,xxu0} and high speed for optical coherent
control.\cite{Press,Kim} Recently, much progress has been made using dot spin
qubits to satisfy the DiVincenzo criteria for quantum
computation,\cite{DiVincenzo} such as spin
initialization,\cite{Atature,Emary,XXu} the coherent manipulation of electron
spins,\cite{Press,Kim2} and fast spin nondestructive measurement.\cite{Kim3}

A fundamental element in quantum computation is the entangling two-qubit
quantum gate. A number of theoretical protocols for two-qubit gate have been
proposed, including through optical RKKY interaction,\cite{Piermarocchi}
Coulomb or tunnelling interactions between excited state in neighbouring
dots,\cite{Calarco,Nazir,Gauger,Emary2} long-range coupling through
waveguide-cavity system,\cite{Clark} and phonon-assisted Zeno
effect.\cite{kjxu} These schemes have yet to be demonstrated experimentally, however.

Recently, ultrafast optical entanglement control utilizing the \textit{ground
state} conduction \textit{electron tunneling} between two quantum-dot spins
was experimentally realized.\cite{D.kim} In this paper, we investigate an
approach that takes advantage of the same type of vertically stacked
self-assembled quantum dots in order to entangle the two conduction electron
spins in the dots, using the \textit{excited} valence \textit{hole tunneling}
as a means to couple the two electron spins. By adjusting the voltage of the
Schottky diode which houses the dots so that the two hole levels line up when
one of the two electrons are optically excited into a trion, we avoid the
tunnel coupling of the two electron spin qubits when there is no optical
excitation. This method enables simpler single qubit operations. The key
physics to accomplish a controlled phase gate is an optical rotation of only
one basis state of the two spins to change its sign, utilizing Pauli blocking
to prevent the unwanted transformation of the remaining three basis states.
Our computed results indicate that by using three laser pulses a controlled
phase gate with a fidelity exceeding 90\% can be implemented on a time scale
as short as 10 ps.

\section{The Basic model\ }

The sample under study contains two vertically coupled SAQDs embedded in a
Schottky diode structure (Fig.~1).\cite{ga1,ga2,ga3} The two vertically
stacked self-assembled InAs/GaAs QDs are separated by a thin GaAs barrier such
that the electrons or holes can tunnel between the two dots. The two QDs have
different thicknesses so that they have different optical transition energies.
As a result they can be optically addressed separately with resonant laser
frequencies. The nominal height of dot $1$, $h_{1}$, is greater than that of
dot $2$, $h_{2}$, so that dot $1$ exhibits the lower transition energy. This
allows the hole levels to be brought into resonance by adjusting the Schottky
diode voltage $V$, while the electron level of dot $2$ is shifted to a higher
energy than that of dot $1$. This is a preliminary step and not part of the
quantum information processing.

Fig.~2 shows the electron spin states and the lowest trion levels for each
quantum dot.\cite{XXu,ywu,Berezovsky} The qubit states are $\left\vert
\uparrow\right\rangle $ and $\left\vert \downarrow\right\rangle $, parallel or
anti-parallel to the $x$ axis (the growth and optical axis). The interband
transition is to the trion state, consisting of two electrons in a singlet
state and a heavy hole. The two trion levels are $\left\vert \uparrow
\downarrow\Uparrow\right\rangle =\frac{1}{\sqrt{2}}\left(  \left\vert
\uparrow\downarrow\right\rangle -\left\vert \downarrow\uparrow\right\rangle
\right)  \left\vert \Uparrow\right\rangle $ and $\left\vert \uparrow
\downarrow\Downarrow\right\rangle =\frac{1}{\sqrt{2}}\left(  \left\vert
\uparrow\downarrow\right\rangle -\left\vert \downarrow\uparrow\right\rangle
\right)  \left\vert \Downarrow\right\rangle $, where $\left\vert
\Uparrow\right\rangle =\left\vert \frac{3}{2}\text{, }\frac{3}{2}\right\rangle
$ and $\left\vert \Downarrow\right\rangle =\left\vert \frac{3}{2}%
\text{,}-\frac{3}{2}\right\rangle $ denote heavy hole states with spin $3/2$
and $-3/2$ components along $x$. Optical selection rules dictate that the
$\sigma^{+}$ polarization could coupled the transition from $\left\vert
\uparrow\right\rangle $ to $\left\vert \uparrow\downarrow\Uparrow\right\rangle
$, and the $\sigma^{-}$ polarization from $\left\vert \downarrow\right\rangle
$ to $\left\vert \downarrow\uparrow\Downarrow\right\rangle $. Here, we have
neglected hole mixing and assumed that the in-plane magnetic field is
zero---conditions to be relaxed later.

\section{Implementation of a two-qubit phase gate}

Among the entangling two-qubit quantum gates, one which is well suited for
realizing with atom-like systems such as quantum dots is the phase gate. The
ideal phase gate aims at a phase change in the basis state $\left\vert
\uparrow\right\rangle _{1}\left\vert \downarrow\right\rangle _{2}$ without
affecting the phases of the other three states. It should also preserve the
phase coherence of a superposition of the four QD spin states. This operation
is a unitary transformation:%
\begin{align}
\left\vert \downarrow\right\rangle _{1}\left\vert \downarrow\right\rangle
_{2}  &  \rightarrow\text{ }\left\vert \downarrow\right\rangle _{1}\left\vert
\downarrow\right\rangle _{2}\text{,}\nonumber\\
\left\vert \downarrow\right\rangle _{1}\left\vert \uparrow\right\rangle _{2}
&  \rightarrow\text{ }\left\vert \downarrow\right\rangle _{1}\left\vert
\uparrow\right\rangle _{2}\text{,}\nonumber\\
\left\vert \uparrow\right\rangle _{1}\left\vert \downarrow\right\rangle _{2}
&  \rightarrow-\left\vert \uparrow\right\rangle _{1}\left\vert \downarrow
\right\rangle _{2}\text{,}\nonumber\\
\left\vert \uparrow\right\rangle _{1}\left\vert \uparrow\right\rangle _{2}  &
\rightarrow\text{ }\left\vert \uparrow\right\rangle _{1}\left\vert
\uparrow\right\rangle _{2}\text{.}%
\end{align}
We use the convention that the vertical arrows in the first and second kets
are, respectively, the directions of the spins in dot $1$ and dot $2$.

To obtain the two-qubit phase gate, we first use a $\sigma^{-}$\ polarized
$\pi$ pulse to excite the dot $1$ spin down state to the trion state. After
waiting for a time interval allowing the hole of the trion state to tunnel
from dot $1$ to dot $2$, we apply a $\sigma^{-}$ $2\pi$ laser pulse to rotate
the dot $2$ spin down state via its trion state. The Pauli blocking given by
the tunneling hole guarantees that only the $\left\vert \uparrow\right\rangle
_{1}\left\vert \downarrow\right\rangle _{2}$ state of the two spins undergoes
$2\pi$ Rabi rotation, and acquires the $-1$ factor, thus realizing a
conditional phase gate. After allowing the hole tunnels back to QD $1$, we use
another $\pi$ laser pulse to de-excite the trion state in dot 1. The system
will return to its original state with a controlled phase shift. Fig.~3 shows
the gate operation process and dynamics of the system under different initial
states. The detailed gate operation process are as follows:

\begin{enumerate}
\item[(i)] An ultrafast $\sigma^{-}$\ polarized $\pi$ pulse (marked
$\Omega_{1}^{\pi}\left(  t\right)  $) is applied to excite the dot $1$ spin
down state to the trion state, thus, the state $\left\vert \downarrow
\right\rangle _{1}\left\vert m\right\rangle _{2}$ ($m=\downarrow,$ $\uparrow$)
to $-i\left\vert \downarrow\uparrow\Downarrow\right\rangle _{1}\left\vert
m\right\rangle _{2}$. (For details see Appendix~\ref{sec_appendix A}.)

\item[(ii)] We utilize the free evolution of the tunnel process by the
Hamiltonian (in the subspace of two tunneling states $\left[  \left\vert
\downarrow\uparrow\Downarrow\right\rangle _{1}\left\vert m\right\rangle
_{2}\text{, }\left\vert \downarrow\uparrow\right\rangle _{1}\left\vert
\Downarrow m\right\rangle _{2}\right]  $),%
\begin{equation}
H_{t}=\left(
\begin{array}
[c]{cc}%
0 & \tau\\
\tau & 0
\end{array}
\right)  \text{,}%
\end{equation}
where $\tau$ is the hole tunneling rate between the two dots. After a precise
time $t_{1}=T_{0}=\pi/(2\tau)$, the state $-i\left\vert \downarrow
\uparrow\Downarrow\right\rangle _{1}\left\vert m\right\rangle _{2}$ will
evolve to $-\left\vert \downarrow\uparrow\right\rangle _{1}\left\vert
\Downarrow m\right\rangle _{2}$. In the two steps of the process, the states
$\left\vert \uparrow\right\rangle _{1}\left\vert \uparrow\right\rangle _{2}$
and $\left\vert \uparrow\right\rangle _{1}\left\vert \downarrow\right\rangle
_{2}$ are not affected by the laser pulse.

\item[(iii)] An ultrafast $\sigma^{-}$\ polarized $2\pi$ pulse (marked
$\Omega_{2}^{2\pi}\left(  t-T_{0}\right)  $) rotates the single trion state in
dot $2$. The state $\left\vert \uparrow\right\rangle _{1}\left\vert
\downarrow\right\rangle _{2}$ makes a complete $2\pi$ rotation through the
state $\left\vert \uparrow\right\rangle _{1}\left\vert \downarrow
\uparrow\Downarrow\right\rangle _{2}$ and thereby acquires an extra $\pi$
phase shift,%
\begin{equation}
\left\vert \uparrow\right\rangle _{1}\left\vert \downarrow\right\rangle
_{2}\rightarrow-\left\vert \uparrow\right\rangle _{1}\left\vert \downarrow
\right\rangle _{2}\text{.}%
\end{equation}
The transition from the state $\left\vert \downarrow\uparrow\right\rangle
_{1}\left\vert \Downarrow\downarrow\right\rangle _{2}$ to $|\downarrow
\uparrow\rangle_{1}|\Downarrow\downarrow,\uparrow\Downarrow\rangle_{2}$ and
back is forbidden by the Pauli exclusion principle. Ideally, the same pulse
does not cause the transition between $\left\vert \downarrow\uparrow
\right\rangle _{1}\left\vert \Downarrow\uparrow\right\rangle _{2}$ and
$\left\vert \downarrow\uparrow\right\rangle _{1}\left\vert \phi\right\rangle
_{2}$ (where $\phi$ denotes the vacuum state) because the excited energy
difference between the exciton and trion $\Delta$ makes this transition to be
off resonance. We will consider the possible deviation from the ideal process later.

\item[(iv)] We utilize the hole tunneling again. After time $t_{2}=T_{0}%
=\pi/(2\tau)$, the state $-\left\vert \downarrow\uparrow\right\rangle
_{1}\left\vert \Downarrow m\right\rangle _{2}$ tunnels back to $i\left\vert
\downarrow\uparrow\Downarrow\right\rangle _{1}\left\vert m\right\rangle _{2}$.

\item[(v)] Another ultrafast $\sigma^{-}$\ polarized $\pi$ pulse (marked as
$\Omega_{1}^{\pi}\left(  t-2T_{0}\right)  $) de-excites the trion state in dot
1, and thus the state $i\left\vert \downarrow\uparrow\Downarrow\right\rangle
_{1}\left\vert m\right\rangle _{2}$ back to $\left\vert \downarrow
\right\rangle _{1}\left\vert m\right\rangle _{2}$. The state $\left\vert
\downarrow\right\rangle _{1}\left\vert m\right\rangle _{2}$ is unchanged by
the entire sequence of operations because it is equivalent to two complete
state rotations.
\end{enumerate}

Table \ref{table1} summarizes the state evolution, in which the first row
numbers the operation steps, the second row shows the time sequence of the
pulse, and each subsequent row shows how the evolution of a basis state in the
first column transforms in time.
\begin{table}[tbp]\centering
\caption{The operation steps, pulse sequence, and state evolution}
\begin{tabular}
[c]{cccccc}\hline\hline
Operation step & (i) & (ii) & (iii) & (iv) & (v)\\
Pulse & $\Omega_{1}^{\pi}\left(  t\right)  $ & $t_{1}$ & $\Omega_{2}^{2\pi
}\left(  t-T_{0}\right)  $ & $t_{2}$ & $\Omega_{1}^{\pi}\left(  t-2T_{0}%
\right)  $\\\hline
$\left\vert \downarrow\right\rangle _{1}\left\vert \downarrow\right\rangle
_{2}$ & $-i\left\vert \downarrow\uparrow\Downarrow\right\rangle _{1}\left\vert
\downarrow\right\rangle _{2}$ &  $ -\left\vert \downarrow\uparrow\right\rangle
_{1}\left\vert \Downarrow\downarrow\right\rangle _{2}$ & $-\left\vert
\downarrow\uparrow\right\rangle _{1}\left\vert \Downarrow\downarrow
\right\rangle _{2}$ &  $i\left\vert \downarrow\uparrow\Downarrow\right\rangle
_{1}\left\vert \downarrow\right\rangle _{2}$ & $\left\vert \downarrow
\right\rangle _{1}\left\vert \downarrow\right\rangle _{2}$\\
$\left\vert \downarrow\right\rangle _{1}\left\vert \uparrow\right\rangle _{2}$
& $-i\left\vert \downarrow\uparrow\Downarrow\right\rangle _{1}\left\vert
\uparrow\right\rangle _{2}$ & $-\left\vert \downarrow\uparrow\right\rangle
_{1}\left\vert \Downarrow\uparrow\right\rangle _{2}$ & $-\left\vert
\downarrow\uparrow\right\rangle _{1}\left\vert \Downarrow\uparrow\right\rangle
_{2}$ & $i\left\vert \downarrow\uparrow\Downarrow\right\rangle _{1}\left\vert
\uparrow\right\rangle _{2}$ & $\left\vert \downarrow\right\rangle
_{1}\left\vert \uparrow\right\rangle _{2}$\\
$\left\vert \uparrow\right\rangle _{1}\left\vert \downarrow\right\rangle _{2}$
& $\left\vert \uparrow\right\rangle _{1}\left\vert \downarrow\right\rangle
_{2}$ & $\left\vert \uparrow\right\rangle _{1}\left\vert \downarrow
\right\rangle _{2}$ & $-\left\vert \uparrow\right\rangle _{1}\left\vert
\downarrow\right\rangle _{2}$ & $-\left\vert \uparrow\right\rangle
_{1}\left\vert \downarrow\right\rangle _{2}$ & $-\left\vert \uparrow
\right\rangle _{1}\left\vert \downarrow\right\rangle _{2}$\\
$\left\vert \uparrow\right\rangle _{1}\left\vert \uparrow\right\rangle _{2}$ &
$\left\vert \uparrow\right\rangle _{1}\left\vert \uparrow\right\rangle _{2}$ &
$\left\vert \uparrow\right\rangle _{1}\left\vert \uparrow\right\rangle _{2}$ &
$\left\vert \uparrow\right\rangle _{1}\left\vert \uparrow\right\rangle _{2}$ &
$\left\vert \uparrow\right\rangle _{1}\left\vert \uparrow\right\rangle _{2}$ &
$\left\vert \uparrow\right\rangle _{1}\left\vert \uparrow\right\rangle _{2}%
$\\\hline\hline
\end{tabular}
\label{table1}
\end{table}

The alternate expression for the phase gate in the basis $\left[  \left\vert
-z\right\rangle _{1}\left\vert -z\right\rangle _{2}\text{, }\left\vert
-z\right\rangle _{1}\left\vert +z\right\rangle _{2}\text{, }\left\vert
+z\right\rangle _{1}\left\vert -z\right\rangle _{2}\text{, }\left\vert
+z\right\rangle _{1}\left\vert +z\right\rangle _{2}\right]  $ is $\frac{1}%
{2}\left[
\begin{array}
[c]{cccc}%
1 & 1 & -1 & 1\\
1 & 1 & 1 & -1\\
-1 & 1 & 1 & 1\\
1 & -1 & 1 & 1
\end{array}
\right]  $, \newline and in the basis $\left[  \left\vert \downarrow
\right\rangle _{1}\left\vert -z\right\rangle _{2}\text{, }\left\vert
\downarrow\right\rangle _{1}\left\vert +z\right\rangle _{2}\text{, }\left\vert
\uparrow\right\rangle _{1}\left\vert -z\right\rangle _{2}\text{, }\left\vert
\uparrow\right\rangle _{1}\left\vert +z\right\rangle _{2}\right]  $ is
$\left[
\begin{array}
[c]{cccc}%
1 & 0 & 0 & 0\\
0 & 1 & 0 & 0\\
0 & 0 & 0 & 1\\
0 & 0 & 1 & 0
\end{array}
\right]  $, where $\left\vert \pm z\right\rangle $ are electron spin states
aligned in the $z$-direction. These expression shows a phase gate in
combination with single-qubit rotations being equivalent to an entanglement
gate or the controlled-not gate.

\section{Hole mixing and unintended dynamics}

For error analysis of the gate operation, we now include the effects of hole
mixing and unintended dynamics in our analysis. From the Luttinger
Hamiltonian,\cite{Luttinger,Broido} the top four states of the valence hole,
rather than the \textquotedblleft bare\textquotedblright\ heavy-hole states
$\left\vert \frac{3}{2}\text{, }\pm\frac{3}{2}\right\rangle $, are mixed by
confinement,
\begin{equation}
h_{\pm}^{\dagger}|\phi\rangle=\cos\theta_{m}\left\vert \frac{3}{2}\text{, }%
\pm\frac{3}{2}\right\rangle -\sin\theta_{m}e^{\mp i\phi_{m}}\left\vert
\frac{3}{2}\text{, }\mp\frac{1}{2}\right\rangle \text{,}%
\end{equation}
where $\left\vert \frac{3}{2}\text{, }\mp\frac{1}{2}\right\rangle $ are
light-hole states aligned in the $x$-direction, and $\theta_{m}$ and $\phi
_{m}$ are mixing angles. With the mixing, the light-matter interaction
Hamiltonian for $\sigma^{-}$\ polarized light with Rabi frequency
$\Omega\left(  t\right)  $ becomes%

\begin{equation}
H_{-}=\frac{\Omega\left(  t\right)  }{2}\sum_{i=1,2}\left(  \cos\theta
_{m}e_{i\uparrow}^{\dagger}h_{i-}^{\dagger}-\sqrt{1/3}\sin\theta_{m}%
e^{-i\phi_{m}}e_{i\downarrow}^{\dagger}h_{i+}^{\dagger}\right)  +H.c.\text{,}%
\end{equation}
here $i=1$, $2$ denote the two quantum dots. A $\sigma^{+}$\ polarized laser
pulse with Rabi frequency $\Omega\left(  t\right)  $ has the Hamiltonian%

\begin{equation}
H_{+}=\frac{\Omega\left(  t\right)  }{2}\sum_{i=1,2}\left(  \cos\theta
_{m}e_{i\downarrow}^{\dagger}h_{i+}^{\dagger}-\sqrt{1/3}\sin\theta_{m}%
e^{i\phi_{m}}e_{i\uparrow}^{\dagger}h_{i-}^{\dagger}\right)  +H.c.\text{,}%
\end{equation}
where the factor of $\sqrt{1/3}$ in the second term comes from the different
weights of in-plane components of the valenceband wave functions. However, by
adjusting the polarizations of the laser, one may establish the actual axis
about which the laser pulse will rotate the state.\cite{Emary3} Instead of
$\sigma^{-}$\ polarized pulse, the new one has the polarization
\begin{equation}
\sigma=\left(  1-2/3\sin^{2}\theta_{m}\right)  ^{-1/2}\left(  \cos\theta
_{m}\sigma^{-}+\sqrt{1/3}\sin\theta_{m}e^{-i\phi_{m}}\sigma^{+}\right)
\text{,}%
\end{equation}
and Rabi frequency $\Omega\left(  t\right)  $, the Hamiltonian can be written
as%
\begin{equation}
H_{\sigma}=\sum_{i=1,2}\frac{\Omega_{\text{eff}}}{2}e_{i\uparrow}^{\dagger
}h_{i-}^{\dagger}+H.c.\text{,}%
\end{equation}
with the effective Rabi frequency
\begin{equation}
\Omega_{\text{eff}}=\Omega\left(  t\right)  \left(  1-2/3\sin^{2}\theta
_{m}\right)  ^{1/2}\text{.}%
\end{equation}
Thus, $\theta_{m}$ and $\phi_{m}$ can be obtained by data fitting after
measuring the effective Rabi frequencies of laser pulses with different
polarizations. The effect of hole mixing is to slightly decrease the effective
Rabi frequency $\Omega_{\text{eff}}$, and this can simply be compensated by a
proportionate increase in $\Omega\left(  t\right)  $. Use of this new
polarization therefore allows us to account for the effects of hole mixing and
proceed directly as outlined in the previous sections.

In order to avoid the unintended excited transition $\left\vert \downarrow
\uparrow\right\rangle _{1}\left\vert \Downarrow\uparrow\right\rangle
_{2}\rightarrow\left\vert \downarrow\uparrow\right\rangle _{1}\left\vert
\phi\right\rangle _{2}$ caused by the ultrafast laser pulse $\Omega_{2}^{2\pi
}\left(  t\right)  $, we propose a remedy by pulse shaping.\cite{chenpc}
Instead of a $2\pi$ single pulse $\Omega_{2}^{2\pi}\left(  t\right)  $, we use
a combination of two phase-locked pulses of $\sigma^{-}$ polarization,
resonant respectively with the transitions $\left\vert \uparrow\right\rangle
_{1}\left\vert \downarrow\right\rangle _{2}\rightarrow\left\vert
\uparrow\right\rangle _{1}\left\vert \downarrow\uparrow\Downarrow\right\rangle
_{2}$ and $\left\vert \downarrow\uparrow\right\rangle _{1}\left\vert
\Downarrow\uparrow\right\rangle _{2}\rightarrow\left\vert \downarrow
\uparrow\right\rangle _{1}\left\vert \phi\right\rangle _{2}$,%
\begin{equation}
\Omega_{2}\left(  t\right)  =\Omega_{2}^{0}\left(  \exp\left[  -(t/s)^{2}%
-i\epsilon_{2}t\right]  -\exp\left[  -(t/s_{1})^{2}-i(\epsilon_{2}%
+\Delta)t\right]  \right)  \text{,}%
\end{equation}
where $\epsilon_{2}$ ($\epsilon_{2}+\Delta$) is the trion (exciton) excited
energy of quantum dot 2. We choose the parameters satisfying the conditions
\begin{align}
\Omega_{2}^{0}(s-s_{1}\exp\left[  -(\Delta s_{1}/2)^{2}\right]  ) &
=\sqrt{\pi}\text{,}\nonumber\\
s_{1}-s\exp\left[  -(\Delta s/2)^{2}\right]   &  =0\text{,}%
\end{align}
\ so that the pulse can produces a $2\pi$ rotation for trion resonance
$\left\vert \uparrow\right\rangle _{1}\left\vert \downarrow\right\rangle
_{2}\rightarrow\left\vert \uparrow\right\rangle _{1}\left\vert \downarrow
\uparrow\Downarrow\right\rangle _{2}$ and brings the exciton pseudospin to the
original state $\left\vert \downarrow\uparrow\right\rangle _{1}\left\vert
\Downarrow\uparrow\right\rangle _{2}$.

Other sources of error are
the spontaneous emission and the laser intensity dependent dephasing as a function of temperature.
The dominant spontaneous emission path is the direct
recombination exciton pair $e_{\uparrow}^{+}h_{\Downarrow}^{+}$ in
the states $\left\vert \downarrow\uparrow\Downarrow\right\rangle _{1}\left\vert
m\right\rangle _{2}$ ($m=\downarrow,$ $\uparrow$), $\left\vert \downarrow
\uparrow\right\rangle _{1}\left\vert \Downarrow\uparrow\right\rangle _{2}$,
and $\left\vert \uparrow\right\rangle _{1}\left\vert \downarrow\uparrow
\Downarrow\right\rangle _{2}$.
Experiments \cite{steel,Ramsay1} showed laser intensity dependence of the exciton Rabi oscillations. The temperature independent effect \cite{steel} is unimportant for control.  We examine the temperature dependent effect, \cite{Ramsay1} which gave
the additional rate of exciton pure dephasing
$\Gamma_{2}\approx AT\Omega^{2}\left(  t\right)  $ due to the
exciton-phonon interaction,
where $A$ is a constant, $T$ is the temperature, and $\Omega$ is the average Rabi frequency of each pulse.
We assess the effects through the
numerical integration of the master equation for the system in the Lindblad
form.\cite{kjxu,Lindblad} Experiments have shown that lifetime of the exciton
is of the order of $t_{e}=1$ ns.\cite{Cortez,Bardot} We choose the laser
pulses, $\Omega_{1}^{\pi}\left(  t\right)  =\frac{\sqrt{\pi}}{2s}\exp\left[
-t^{2}/s^{2}-i\epsilon_{1}t\right]  $, here $\epsilon_{1}$ is the trion
excited energy of QD 1, $\Omega_{2}\left(  t\right)  $ is defined by
Eqs.~(10-11), with $s=0.2$ ps, $\Delta=4$ meV,\cite{Stinaff} and the tunneling
$\tau=2$ meV,\cite{ga3} $T_{0}=\pi/(2\tau)=3.27$ ps, $T=1$ K, $A=11$ fs $\cdot K^{-1}$
taken from Ref.~\onlinecite{Ramsay1}.
For the initial state
\begin{equation}
\left\vert \Psi^{0}\right\rangle =\frac{1}{2}\left(  \left\vert \downarrow
\right\rangle _{1}\left\vert \downarrow\right\rangle _{2}+\left\vert
\downarrow\right\rangle _{1}\left\vert \uparrow\right\rangle _{2}+\left\vert
\uparrow\right\rangle _{1}\left\vert \downarrow\right\rangle _{2}+\left\vert
\uparrow\right\rangle _{1}\left\vert \uparrow\right\rangle _{2}\right),
\end{equation}
the dynamics of the density matrix elements $\rho\left(  \left\vert
\downarrow\right\rangle _{1}\left\langle \downarrow\right\vert \otimes
\left\vert \downarrow\right\rangle _{2}\left\langle \downarrow\right\vert
\right)  $, $\rho\left(  \left\vert \downarrow\right\rangle _{1}\left\langle
\downarrow\right\vert \otimes\left\vert \uparrow\right\rangle _{2}\left\langle
\uparrow\right\vert \right)  $, $\rho\left(  \left\vert \uparrow\right\rangle
_{1}\left\langle \uparrow\right\vert \otimes\left\vert \downarrow\right\rangle
_{2}\left\langle \downarrow\right\vert \right)  $, $\rho\left(  \left\vert
\uparrow\right\rangle _{1}\left\langle \uparrow\right\vert \otimes\left\vert
\downarrow\right\rangle _{2}\left\langle \uparrow\right\vert \right)  $ are
shown in Fig.~4. The figure shows the key
feature of the phase gate that,
after the $\Omega_{2}(t)$ pulse, the joint two-qubit coherence (or
off-diagonal element) $\rho(\left\vert \uparrow\right\rangle _{1}\left\langle
\uparrow\right\vert \otimes\left\vert \downarrow\right\rangle _{2}\left\langle
\uparrow\right\vert )$ gains a minus sign. Simultaneously, the joint
\textquotedblleft up-down\textquotedblright\ population, $\rho(\left\vert
\uparrow\right\rangle _{1}\left\langle \uparrow\right\vert \otimes\left\vert
\downarrow\right\rangle _{2}\left\langle \downarrow\right\vert )$, returns to
its original value after the $2\pi$ rotation. Similarly, both the
\textquotedblleft down-down\textquotedblright\ and \textquotedblleft
down-up\textquotedblright\ populations, $\rho(\left\vert \downarrow
\right\rangle _{1}\left\langle \downarrow\right\vert \otimes\left\vert
\downarrow\right\rangle _{2}\left\langle \downarrow\right\vert ),\rho
(\left\vert \downarrow\right\rangle _{1}\left\langle \downarrow\right\vert
\otimes\left\vert \uparrow\right\rangle _{2}\left\langle \uparrow\right\vert
)$, return to their initial values with visible errors after the pulses
$\Omega_{1}^{\pi}\left(  t\right)  $ and $\Omega_{1}^{\pi}\left(
t-2T_{0}\right)  $. The \textquotedblleft up-up\textquotedblright%
\ populations, $\rho(\left\vert \uparrow\right\rangle _{1}\left\langle
\uparrow\right\vert \otimes\left\vert \uparrow\right\rangle _{2}\left\langle
\uparrow\right\vert )$, not shown in the figure, is unchanged by the gate
operation. The figure also shows that the total time of implementing the phase
gate is $T_{g}\approx7$ ps. We calculate the fidelity of the phase gate
$F=0.956$. Besides the spontaneous emission, the error of tunneling time also
decreases the fidelity of the gate. For a $10\%$ error in the tunneling time,
the fidelity is further reduced to $F=0.94$.

Fig.~5 shows the notable effect of the temperature and intensity dependent exciton dephasing on the  fidelity of the phase gate as a function of the pulse duration $s$ and for
the temperatures $T=0$ -- 4 K, with the parameters in the caption.

\section{Compatibility with single-qubit rotation}

In addition to a two-qubit phase gate, single-qubit rotations are required to
demonstrate universal quantum computing. In order to make our operation
compatible with the single qubit rotation schemes in Refs.~\onlinecite{Emary3}
and~\onlinecite{pochung}, a static magnetic field $B$ is required in the $z$
direction (perpendicular to the growth direction $x$). The single-qubit
operations can be performed by using off-resonance Raman processes through the
virtual excitation of an exciton of a single dot. The QDs can be optically
addressed separately with resonant laser frequencies.

The application of the magnetic field renders the qubits in the phase gate no
longer energy eigenstates. Consequently, the effect of the magnetic field on
the fidelity of the gate has to be examined. The qubit states, $\left\vert
\uparrow\right\rangle =\frac{1}{\sqrt{2}}\left(  \left\vert +z\right\rangle
+\left\vert -z\right\rangle \right)  $ and $\left\vert \downarrow\right\rangle
=\frac{1}{\sqrt{2}}\left(  \left\vert +z\right\rangle -\left\vert
-z\right\rangle \right)  $, now precess about the $z$ axis at the Larmor
frequency $g_{z}^{e}\mu_{B}B/\hbar$, where $g_{z}^{e}$ is the effective
electron in-plane g factor and $\mu_{B}$ is the Bohr magneton. The excited
hole states are $\left\vert \Uparrow\right\rangle $ and $\left\vert
\Downarrow\right\rangle $, precess at the frequency $g_{z}^{h}\mu_{B}B/\hbar$,
where $g_{z}^{h}$ is the effective hole in-plane g factor. To find the
magnetic field effect on the phase gate fidelity, we calculate the dynamics of
the system using the measured values \cite{XXu} $g_{z}^{e}=0.48$, $g_{z}%
^{h}=0.31$ and the Gaussian pulses, $\Omega_{1}^{\pi}\left(  t\right)
=\frac{\sqrt{\pi}}{2s}\exp\left[  -t^{2}/s^{2}-i\epsilon_{1}t\right]  $,
$\Omega_{2}\left(  t\right)  $ defined by Eqs.~(10-11), $\Delta=4$ meV, the
tunneling $\tau=2$ meV, $T_{0}=\pi/(2\tau)=3.27$ ps, the trion lifetime
$t_{e}=1$ ns, $A=11$ fs $\cdot K^{-1}$, and the temperature $T=1$ K. For an initial state $\left\vert \Psi^{0}\right\rangle $ in
Eq.~(12), we plot the phase gate fidelity $F$ as a function of the magnetic
field $B$ and the inverse pulse duration $s^{-1}$ in Fig.~6. We can see that
the fidelity decreases with increasing magnetic field, because the bandwidth
of the gate pulses must be larger than the Zeeman splitting. The controlled
phase gate can be implemented with a fidelity over $0.90$ in the case of
$B=1$ T and $s^{-1}=1$ THz.

\section{Conclusion}

In conclusion, we have proposed a controlled-phase gate for two coupled SAQDs
in the Voigt configuration, which is compatible with the previously designed
single qubit rotations.\cite{Emary3,pochung} The speed of our gate is
essentially limited by the hole tunneling between the two quantum dots. In
Ref.~\onlinecite{D.kim} the tunneling of electron is always coupling the two
spins of quantum dots, which enables simpler two-qubit rotation while more
difficult single qubit operations than our scheme. We have shown that hole
mixing can be simply incorporated into this scheme through a change in laser
polarization. The result shows that we could implement the gate in $10$ ps
range and fidelity over $90\%$. Our proposal therefore offers an accessible
path to the demonstration of ultrafast quantum logic in SAQDs.

\textbf{Acknowledgments: }L. Chen thanks Dr. W. Yang and Prof. Y.J. Gu for
helpful discussions. This research was supported by the U.S. Army Research
Office MURI award W911NF0910406. L. Chen was also supported in part by the
Government of China through CSC Grant No. 2009633075.

\appendix

\section{Excitation of the dot 1 spin down state to the trion state}

\label{sec_appendix A} The coupled SAQDs are illuminated with a $\sigma^{-}$
circularly polarized laser pulse propagating in the $x$ direction. The laser
is tuned such that it could create an exciton in the quantum dot $1$, only if
it is state is $\left\vert \downarrow\right\rangle $. This will not affect the
quantum dot $2$ because an trion in the smaller dot $2$ is tens of
millielectron volt (mev) higher than dot $1$ in energy.\cite{D.kim} In the
subspaces $\left[  \left\vert \downarrow\right\rangle _{1}\left\vert
m\right\rangle _{2}\text{, }\left\vert \downarrow\uparrow\Downarrow
\right\rangle _{1}\left\vert m\right\rangle _{2}\text{, }\left\vert
\downarrow\uparrow\right\rangle _{1}\left\vert \Downarrow m\right\rangle
_{2}\right]  $ ($m=\downarrow,$ $\uparrow$), the Hamiltonian for the QDM under
this laser excitation is%

\begin{equation}
H=\left(
\begin{array}
[c]{ccc}%
0 & \frac{\Omega_{0}}{2} & 0\\
\frac{\Omega_{0}}{2} & 0 & \tau\\
0 & \tau & 0
\end{array}
\right)  \text{,}%
\end{equation}
where the energy is in units of $\hbar$, $\Omega_{0}$ is the Rabbi frequency
of the laser pulse, and $\tau$ is the hole tunneling rate between the two
dots. Transforming the basis set of the excited hole states, one in each dot, to a basis of tunneling eigenstates,
$\left\vert \Phi^{\pm}\right\rangle =2^{-1/2}\left(  \left\vert \downarrow
\uparrow\Downarrow\right\rangle _{1}\left\vert m\right\rangle _{2}%
\pm\left\vert \downarrow\uparrow\right\rangle _{1}\left\vert \Downarrow
m\right\rangle _{2}\right)  $, we have%
\begin{equation}
H=\left(
\begin{array}
[c]{ccc}%
0 & \frac{\Omega_{0}}{2\sqrt{2}} & \frac{\Omega_{0}}{2\sqrt{2}}\\
\frac{\Omega_{0}}{2\sqrt{2}} & -\tau & 0\\
\frac{\Omega_{0}}{2\sqrt{2}} & 0 & \tau
\end{array}
\right)  \text{.}%
\end{equation}
Starting with the initial state $\left\vert \downarrow\right\rangle
_{1}\left\vert m\right\rangle _{2}$, the system evolves at the time $t$ to
\begin{equation}
\left\vert \psi\left(  t\right)  \right\rangle =\frac{1}{\sqrt{2}\left(
\tau^{2}+\Omega_{0}^{2}/4\right)  }\left(
\begin{array}
[c]{c}%
\sqrt{2}\tau^{2}+\sqrt{2}\Omega_{0}^{2}/4\cos\left(  t\sqrt{\tau^{2}%
+\Omega_{0}^{2}/4}\right) \\
-i\Omega_{0}/2\sqrt{\tau^{2}+\Omega_{0}^{2}/4}\sin\left(  t\sqrt{\tau
^{2}+\Omega_{0}^{2}/4}\right)  +\tau\Omega_{0}/2\cos\left(  t\sqrt{\tau
^{2}+\Omega_{0}^{2}/4}\right)  -\tau\Omega_{0}/2\\
-i\Omega_{0}/2\sqrt{\tau^{2}+\Omega_{0}^{2}/4}\sin\left(  t\sqrt{\tau
^{2}+\Omega_{0}^{2}/4}\right)  -\tau\Omega_{0}/2\cos\left(  t\sqrt{\tau
^{2}+\Omega_{0}^{2}/4}\right)  +\tau\Omega_{0}/2
\end{array}
\right)  .
\end{equation}
At time $t_{0}=\pi/(2\sqrt{\tau^{2}+\Omega_{0}^{2}/4})$, the state is
\begin{equation}
\left\vert \psi\left(  t_{1}\right)  \right\rangle =\frac{1}{\sqrt{2}\left(
\tau^{2}+\Omega_{0}^{2}/4\right)  }\left(
\begin{array}
[c]{c}%
\sqrt{2}\tau^{2}\\
-i\Omega_{0}/2\sqrt{\tau^{2}+\Omega_{0}^{2}/4}-\tau\Omega_{0}/2\\
-i\Omega_{0}/2\sqrt{\tau^{2}+\Omega_{0}^{2}/4}+\tau\Omega_{0}/2
\end{array}
\right)  \text{.}%
\end{equation}
In the case $\Omega_{0}\gg\tau$, $t_{0}\approx\pi/\Omega_{0}$, $\left\vert
\psi\left(  t_{1}\right)  \right\rangle \approx\frac{-i}{\sqrt{2}}\left(  \left\vert
\Phi^{+}\right\rangle +\left\vert \Phi^{-}\right\rangle \right)  =-i\left\vert
\downarrow\uparrow\Downarrow\right\rangle _{1}\left\vert m\right\rangle _{2}$.
So for a sufficiently short duration $\pi$ pulse, the equal combination state of the two tunneling eigenstates, $\left\vert \Phi^{+}\right\rangle $ and $\left\vert \Phi^{-}\right\rangle$,  is created which corresponds to a localized trion state $\left\vert \downarrow\uparrow\Downarrow\right \rangle_{1}\left\vert m\right\rangle _{2}$.

\newpage\begin{figure}[ptb]
\includegraphics[scale=1.2, angle=0]{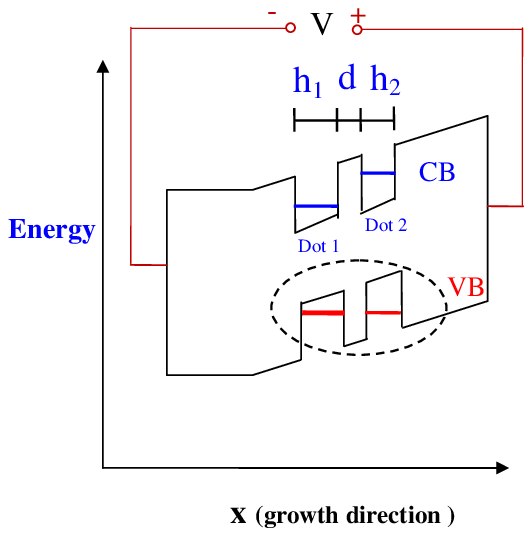}\caption{[Color online]
Schematic diagram of the vertically coupled quantum dot system.\cite{ga1,ga2}
The height of the dot 1 is $h_{1}$, that of dot 2 is $h_{2}$, the interdot
barrier is $d$. The hole levels can be brought into resonance by adjusting the
Schottky diode voltage $V$.}%
\end{figure}

\newpage\begin{figure}[ptb]
\includegraphics[scale=1.4, angle=0]{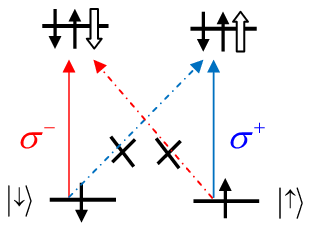}\caption{[Color online] The
level diagram of a charged quantum dot with the one-electron spin states and
the optically allowed transitions to the trion states. The short solid arrows
represent electrons and the short open arrows represent heavy holes. All the
arrows are aligned in the $x$-direction. Long arrows with solid lines indicate
allowed optical transitions with $\sigma^{+}$ and $\sigma^{-}$ denoting two
orthogonal circular polarizations. Long arrows with dotted lines and the
crosses (X) denote optical transitions are forbidden.}%
\end{figure}

\newpage\begin{figure}[ptb]
\includegraphics[scale=0.5, angle=0]{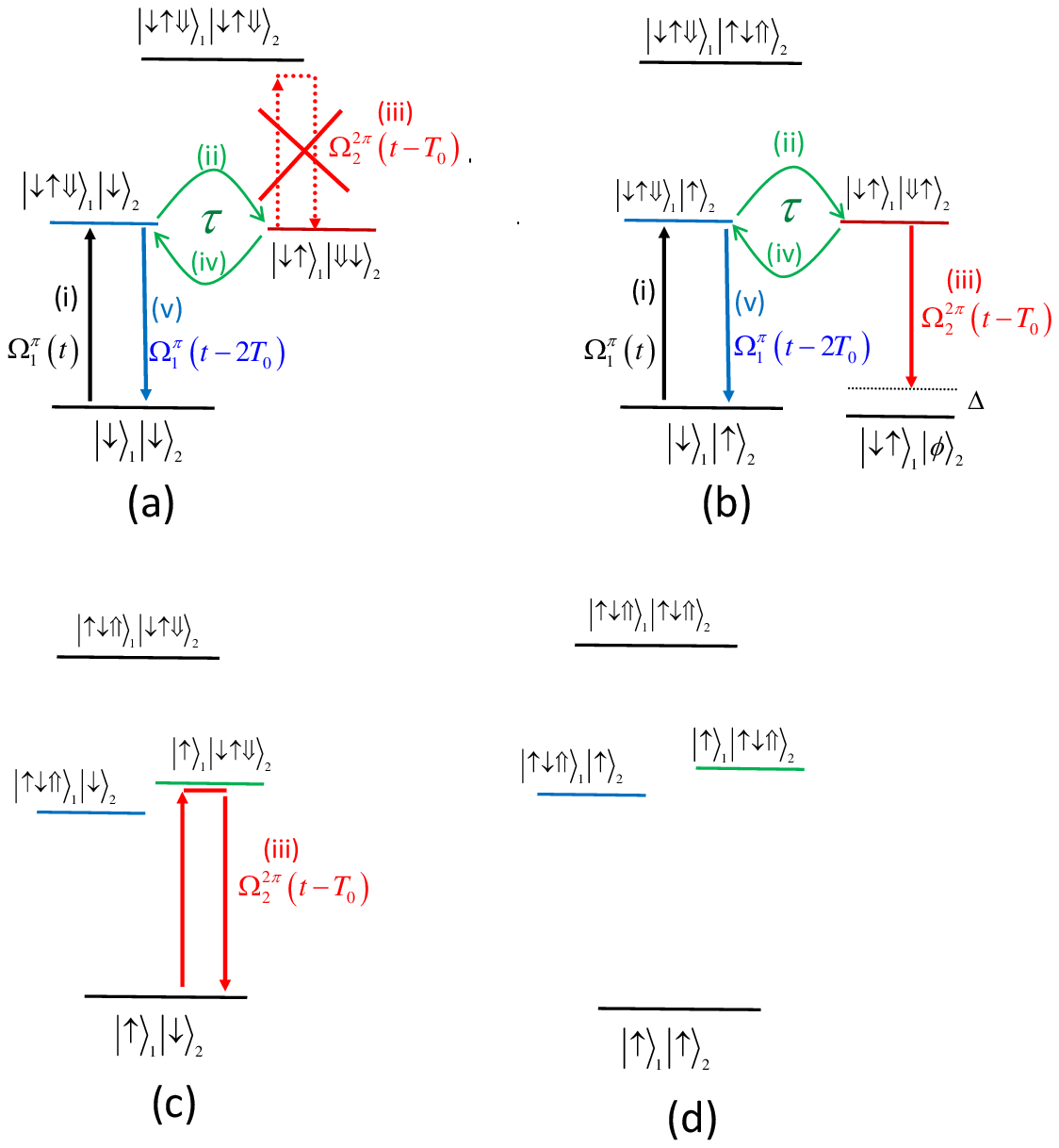}\caption{[Color online] The gate
operation process and dynamics of the system under different initial states.
The labels (i)-(v) correspond to the operation steps. $\tau$ is the hole
tunneling between the two dots. Figure (a)-(c) (i) If the electron in QD $1$
is spin down, the trion state can be excited by the laser pulse $\Omega
_{1}^{\pi}\left(  t\right)  $, thus, the state $\left\vert \downarrow
\right\rangle _{1}\left\vert m\right\rangle _{2}$ ($m=\downarrow,$ $\uparrow$)
to $\left\vert \downarrow\uparrow\Downarrow\right\rangle _{1}\left\vert
m\right\rangle _{2}$. (ii) The hole of the trion in QD 1 tunnels to the QD2,
thus, the state $\left\vert \downarrow\uparrow\Downarrow\right\rangle
_{1}\left\vert m\right\rangle _{2}$ to $\left\vert \downarrow\uparrow
\right\rangle _{1}\left\vert \Downarrow m\right\rangle _{2}$. (iii) The laser
pulse $\Omega_{2}^{2\pi}\left(  t-T_{0}\right)  $ performing a $2\pi$ rotation
between $\left\vert \uparrow\right\rangle _{1}\left\vert \downarrow
\right\rangle _{2}$ and $\left\vert \uparrow\right\rangle _{1}\left\vert
\downarrow\uparrow\Downarrow\right\rangle _{2}$ acquiring the $\pi$ phase
shift, while transition of state $\left\vert \downarrow\uparrow\right\rangle
_{1}\left\vert \Downarrow\downarrow\right\rangle _{2}$ is forbidden because of
Pauli blocking (denoted by the large X ). Ideally, the same pulse does not
cause the transition between $\left\vert \downarrow\uparrow\right\rangle
_{1}\left\vert \Downarrow\uparrow\right\rangle _{2}$ and $\left\vert
\downarrow\uparrow\right\rangle _{1}\left\vert \phi\right\rangle _{2}$ (where
$\phi$ denotes the vacuum state) because excited energy difference between
exciton and trion $\Delta$ makes this transition to be off resonance. (iv) The
state $\left\vert \downarrow\uparrow\right\rangle _{1}\left\vert \Downarrow
m\right\rangle _{2}$\ tunnels back to $\left\vert \downarrow\uparrow
\Downarrow\right\rangle _{1}\left\vert m\right\rangle _{2}$. (v) The laser
pulse $\Omega_{1}^{\pi}\left(  t-2T_{0}\right)  $ de-excite the trion state in dot
1, thus, the state $\left\vert \downarrow\uparrow\Downarrow\right\rangle
_{1}\left\vert m\right\rangle _{2}$ back to $\left\vert \downarrow
\right\rangle _{1}\left\vert m\right\rangle _{2}$. Figure (d) If the electrons
in both QDs are spin up, they are not affected by the $\sigma^{-}$\ polarized
laser pulses.}%
\end{figure}

\newpage\begin{figure}[ptb]
\includegraphics[scale=1.0, angle=0]{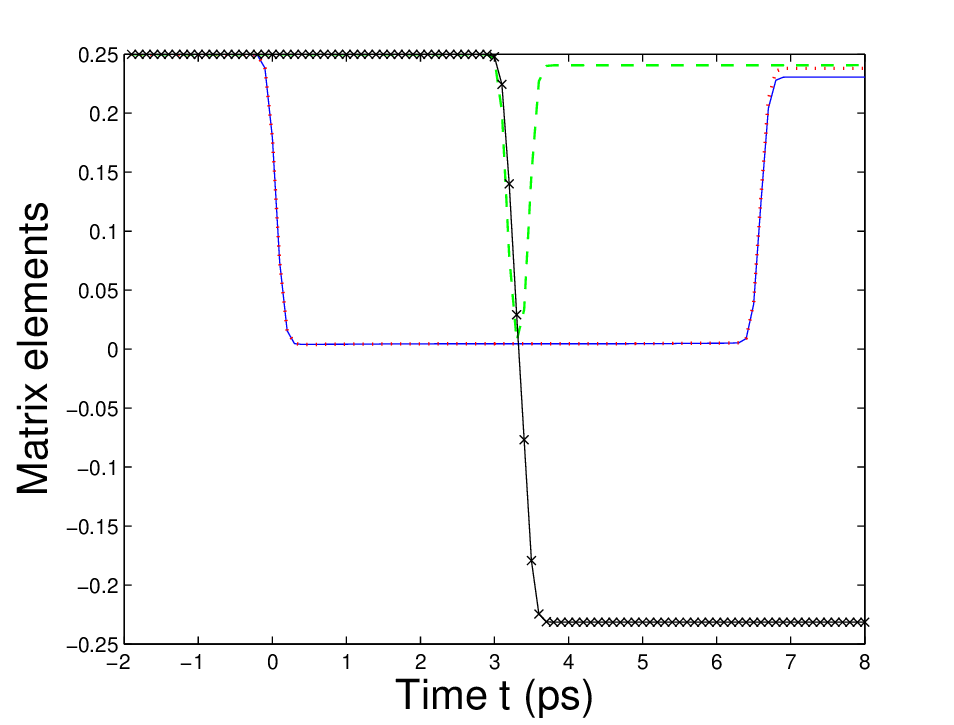}\caption{[Color online]
Dynamical evolution of selected density matrix elements for the initial state
$\left\vert \Psi^{0}\right\rangle $\ in Eqs.~(10) during the gate operation
via numerical simulation. The selected density matrix elements are
$\rho\left(  \left\vert \downarrow\right\rangle _{1}\left\langle
\downarrow\right\vert \otimes\left\vert \downarrow\right\rangle _{2}%
\left\langle \downarrow\right\vert \right)  $ denoted by the solid (blue)
line, $\rho\left(  \left\vert \downarrow\right\rangle _{1}\left\langle
\downarrow\right\vert \otimes\left\vert \uparrow\right\rangle _{2}\left\langle
\uparrow\right\vert \right)  $ the dotted (red) line, $\rho\left(  \left\vert
\uparrow\right\rangle _{1}\left\langle \uparrow\right\vert \otimes\left\vert
\downarrow\right\rangle _{2}\left\langle \downarrow\right\vert \right)  $ the
dashed (green) line, and $\rho\left(  \left\vert \uparrow\right\rangle
_{1}\left\langle \uparrow\right\vert \otimes\left\vert \downarrow\right\rangle
_{2}\left\langle \uparrow\right\vert \right)  $ the $\times$ marked (black)
line.}%
\end{figure}

\newpage\begin{figure}[ptb]
\includegraphics[scale=1.0, angle=0]{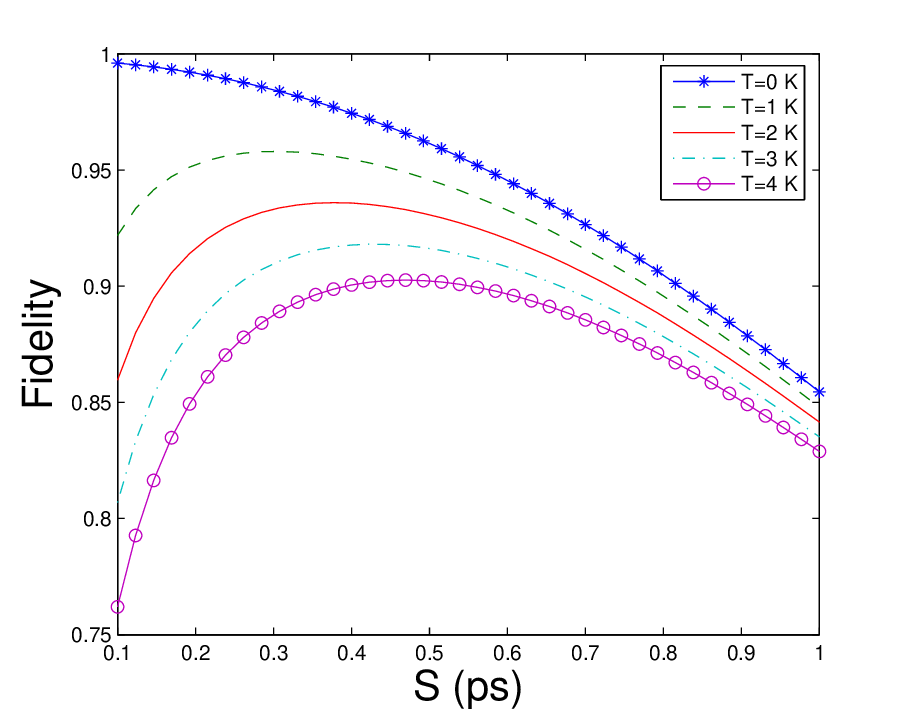}\caption{[Color online] Fidelity of the
phase gate as a function of the pulse duration
$S$ and for
the temperatures $T=0$~--~$4$~K, with the parameters $\Delta=4$ meV,
$g_{z}^{e}=0.48$, $g_{z}^{h}=0.31$, $\tau=2$ meV, $t_{e}=1$ ns, and $A=11$ fs $\cdot K^{-1}$.
\protect{\cite{Ramsay1}} }
\end{figure}

\newpage\begin{figure}[ptb]
\includegraphics[scale=1.0, angle=0]{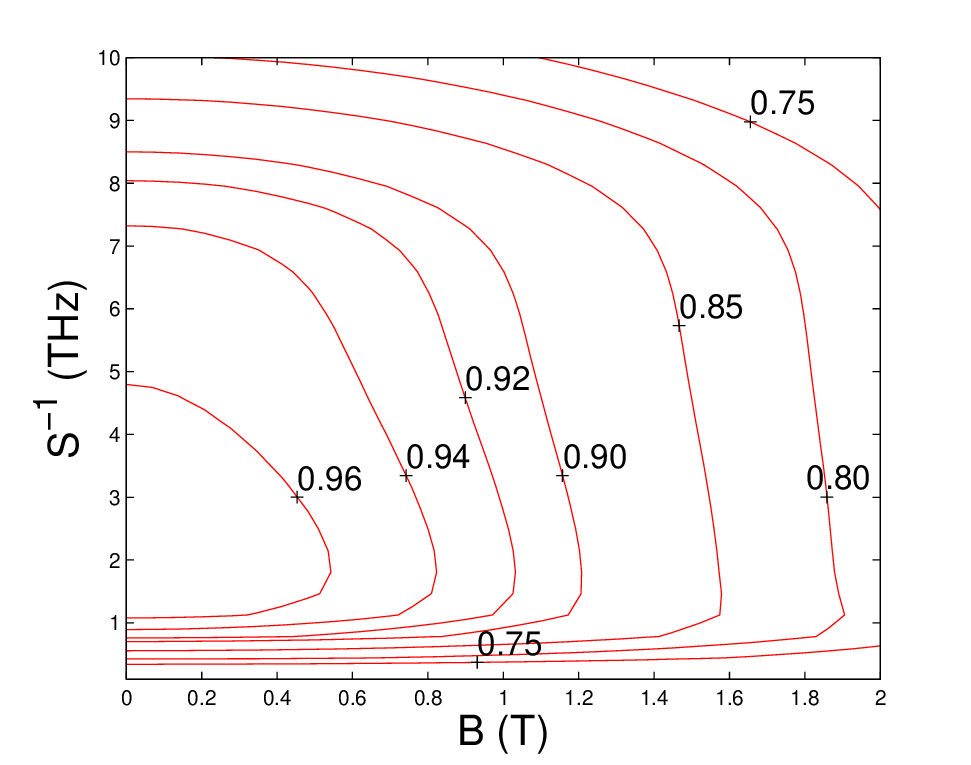}\caption{[Color online] Contour
plot of the phase gate fidelity $F$ for the initial state $\left\vert \Psi
^{0}\right\rangle $ in Eqs.~(12) as a function of the magnetic field $B$ and
the inverse pulse duration $s^{-1}$ with the parameters $\Delta=4$ meV,
$g_{z}^{e}=0.48$, $g_{z}^{h}=0.31$, $\tau=2$ meV, $t_{e}=1$ ns, $T=1$ K, and\protect{\cite{Ramsay1}}  $A=11$ fs $\cdot K^{-1}$.
}
\end{figure}

\end{document}